\documentstyle[amssymb,aps,epsfig,preprint,pre]{revtex}

\begin{document}

\title{Computer simulations of domain growth and phase separation in
  two-dimensional binary immiscible fluids using dissipative particle
  dynamics}

\author{Peter V. Coveney} 
\address{Schlumberger Cambridge Research, 
  High Cross, Madingley Road, Cambridge, CB3 0EL, U.K. \\
  {\tt coveney@cambridge.scr.slb.com}}

\author{Keir E. Novik}
\address{Cavendish Laboratory, Cambridge University, 
  Madingley Road, Cambridge, CB3 0HE, U.K. \\
  {\tt ken21@cam.ac.uk}}

\date{June 27, 1996}
\maketitle


\begin{abstract}
  We investigate the dynamical behavior of binary fluid systems in two
  dimensions using dissipative particle dynamics.  We find that
  following a symmetric quench the domain size $R(t)$ grows with time
  $t$ according to two distinct algebraic laws $R(t) \sim t^{n}$: at
  early times $n = \frac{1}{2}$, while for later times $n =
  \frac{2}{3}$.  Following an asymmetric quench we observe only $n =
  \frac{1}{2}$, and if momentum conservation is violated we see $n =
  \frac{1}{3}$ at early times.  Bubble simulations confirm the
  existence of a finite surface tension and the validity of Laplace's
  law. Our results are compared with similar simulations which have
  been performed previously using molecular dynamics, lattice-gas and
  lattice-Boltzmann automata, and Langevin dynamics. We conclude that
  dissipative particle dynamics is a promising method for simulating
  fluid properties in such systems.
\end{abstract}

\pacs{51.10.+y, 02.70.-c, 64.75.+g}
\narrowtext

\section{Introduction}
\label{sec:Intro}

Growth kinetics in binary immiscible fluids has received much
attention recently.  Phase separation in these systems has been
simulated using a variety of techniques, including cell dynamical
systems without hydrodynamics~\cite{bib:so} and with Oseen tensor
hydrodynamics~\cite{bib:so2}; time-dependent Ginzburg-Landau models
without hydrodynamics~\cite{bib:ctg}, and with
hydrodynamics~\cite{bib:fv,bib:vf,bib:walc,bib:lwac}; as well as
lattice-gas
automata~\cite{bib:bal,bib:rothmanrefs,bib:appert,emerton:domains} and
the related lattice-Boltzmann techniques~\cite{bib:acg,bib:oosyb}.  A
central quantity in the study of growth kinetics is the time-dependent
average domain size $R(t)$. For binary systems in the regime of sharp
domain walls, this follows algebraic growth laws of the form $R(t)
\sim t^{n}$.  In general, for models without hydrodynamic interactions
(that is, when there is no conservation of momentum, as is often
supposed to be the case for binary alloys) the growth exponent is
found to be $n = \frac{1}{3}$, independent of the spatial dimension.
If flow effects are relevant (as is certainly the case for binary
fluids), and the domain size $R$ is greater than the hydrodynamic
length $R_{h} = \nu^2 / \rho \sigma$, where $\nu$ is the kinematic
viscosity, $\rho$ is the density and $\sigma$ is the surface tension
coefficient, then one obtains $n = \frac{2}{3}$ in two spatial
dimensions~\cite{bib:bray}.  In the less commonly observed $R < R_{h}$
regime, two-dimensional lattice-gas automata~\cite{emerton:domains},
molecular dynamics~\cite{bib:vat} and Langevin dynamics
simulations~\cite{bib:walc,bib:lwac} find $n = \frac{1}{2}$;
lattice-Boltzmann studies, by contrast, suggest that $n =
\frac{1}{3}$~\cite{bib:oosyb}.  However, the lattice-Boltzmann method
does not include any thermal fluctuations which a renormalization
group approach shows play a crucial role in causing this exponent to
assume the value of $\frac{1}{2}$~\cite{bib:bray}.  In three
dimensions, for $R < R_{h}$ the growth exponent is $n = \frac{1}{3}$
crossing over to $n = 1$ at late times, with $n = \frac{2}{3}$ again
if $R > R_{h}$~\cite{bib:bray}.

It should be noted that there are still some experimental and
theoretical challenges in unraveling the behavior of systems in which
both the order parameter and the momentum are locally
conserved~\cite{bib:bray}.  Experimentally, for example, it is
difficult if not impossible to study two-dimensional fluid systems. 
As far as numerical studies are concerned, it is important to recognize that
three-dimensional simulations are particularly demanding on all the
aforementioned techniques and so definitive results are harder to come
by than in two dimensions.

The purpose of the present paper is to take a look at domain growth
and phase separation in two-dimensional, binary, immiscible fluids
using a new simulation technique called dissipative particle dynamics
(DPD). The basic features of the method are discussed in
Sec.~\ref{sec:WhyDPD}; here we simply note that it is a temporally
discrete and spatially continuous, microscopic, particulate technique
which yields correct hydrodynamical behavior in the macroscopic limit,
while being easy to extend from two to three dimensions. Comparatively
little work has been published on DPD and, to our knowledge, nothing
at all on its application to phase ordering kinetics.  We shall find
that the method is able to handle domain growth both qualitatively and
quantitatively, yielding the correct scaling exponents and displaying
a surface tension which satisfies Laplace's law.  Although the present
study is confined to the case of two-dimensional systems, we hope to
return in the near future with a second paper dealing with the
three-dimensional case.



\section{Why Dissipative Particle Dynamics?}
\label{sec:WhyDPD}

The motivation for the introduction of dissipative particle dynamics
by Hoogerbrugge and Koelman~\cite{hoogerbrugge:dpd} was to simulate the
behavior of complex fluids. Complex fluids include fluids in which
there are many coexisting length and time scales, and fluids for which
a hydrodynamic description is unknown or does not exist at all.
Examples are multiphase flows, flow in porous media, colloidal
suspensions, microemulsions and polymeric fluids.  Whereas the
traditional continuum-based approach to understanding and modeling
the hydrodynamic behavior of such fluids, based on the formulation and
solution of partial differential equations, has met with rather
limited success, in more recent times new approaches have been
undertaken, relying on a microscopic description of the fluid in
question.  In principle, the most accurate microscopic approach is
based on the use of molecular dynamics (MD) but the computational
expense of attempting this is so severe that until now only a small
number of flow phenomena in simple fluids have been achieved, and even
then these have been restricted to two dimensions.

Lattice-gas automata (LGA) have been used as a numerical technique for
modeling hydrodynamics since their introduction in 1986 by Frisch,
Hasslacher and Pomeau~\cite{bib:fhp} and by
Wolfram~\cite{bib:wolfram}, who showed that one could simulate the
incompressible Navier-Stokes equations for a single component fluid
using discrete Boolean elements on a triangular lattice. In essence,
LGA dynamics is comprised of two elements: at each discrete timestep,
particles first collide at vertices on the lattice, the collisional
outcomes being controlled by local conservation of mass and momentum;
then the particles advect freely to neighboring sites.  Such lattice
gas automaton models are computationally much faster than molecular
dynamics, particularly since the natural time step---the mean free
time between collisions---is several orders of magnitude greater than
that required for MD.  The single phase LGA method was subsequently
generalized by Rothman and Keller~\cite{bib:rk} to permit the
simulation of immiscible fluids, and indeed our present work shares
certain features in common with their model, which has since been
investigated with some degree of
rigor~\cite{bib:rothmanrefs,bib:appert}.  Even more recently, the
technique has been extended to model amphiphilic fluids comprised of
mixtures of oil, water and surfactant~\cite{bib:bce}.

However, the LGA method suffers from some disadvantages: the
underlying lattice leads to the loss of Galilean invariance which,
although negligible for creep flows, does present problems for flow at
finite Reynolds numbers, while the treatment of three-dimensional
fluids is computationally challenging owing to the complexity of the
collisional look-up tables and the necessity of deploying a
four-dimensional face-centered hypercubic
lattice~\cite{bib:rothmanrefs}.  The lack of Galilean invariance
manifests itself by a spurious factor multiplying the inertial term in
the momentum-conserving Navier-Stokes equation.  For a single-phase
lattice gas, this factor can easily be scaled away; for compressible
flow, or for multiphase flow with interfaces, however, the presence of
this factor is more difficult to deal with, although various rather
involved methods have been proposed to remove
it~\cite{bib:dhl,bib:bt}.

Dissipative particle dynamics was introduced with the intention of
capturing the best aspects of MD and LGA; it does away with the
problems arising in the latter owing to the presence of a lattice,
while maintaining the discrete time-stepping element which greatly
accelerates the algorithm compared with MD.  Moreover---and this is
important from a practical perspective---in DPD the extension from two
to three dimensions is very straightforward.  The DPD method involves
the motion of massive particles, which are allowed to move in space
with their positions and momenta described by real numbers.  The mass
and momentum of these particles are conserved, but their energy is
not.  As in LGA, the evolution of the model over one time step takes
place in two substeps which are continually repeated: (i) an
infinitesimally short impulse step, and (ii) a propagation step of
duration $\Delta t$.  Within the impulse substep, the momentum of each
particle (${\mathbf p}_i$) is modified to reflect its interaction with
the other particles.  During the propagation step each particle coasts
with constant velocity, completely ignoring every other particle.  In
mathematical terms, the impulse step is described by
\begin{equation}\label{ImpulseStep}
\Delta {\mathbf p}_i = \sum_{j \ne i}\Omega_{ij} {\mathbf\hat e}_{ij},
\end{equation}
while the propagation step is
\begin{equation}\label{PropagationStep}
\Delta {\mathbf q}_i = {\Delta t \over m} \left( {\mathbf p}_i + 
\Delta {\mathbf p}_i \right) .
\end{equation}
In these equations $m$ is the mass of each particle, ${\mathbf q}_i$
denotes the position of particle $i$, and ${\mathbf\hat e}_{ij}$ is
the unit vector pointing from particle $j$ to particle $i$.
Henceforth, we shall assume for simplicity that all particles carry
unit mass ($m = 1$).  $\Omega_{ij}$ is a scalar giving the momentum
transferred from $j$ to $i$, and in the original model presented by
Hoogerbrugge and Koelman~\cite{hoogerbrugge:dpd} has the form
\begin{equation}\label{OriginalOmega}
  \Omega_{ij} = \cases{ \displaystyle
    {3 \over \pi r_c^2 n} \left( 1 - {r_{ij} \over r_c} \right) \left[
      \Pi_{ij} - \omega \left( {\mathbf p}_i - {\mathbf p}_j \right)
      \cdot {\mathbf\hat e}_{ij} \right]
& if $r_{ij} < r_c$, \cr
0 & if $r_{ij} \ge r_c$, \cr
}
\end{equation}
where $r_{ij} = \left| {\mathbf q}_i - {\mathbf q}_j \right|$ is the
distance between particles $i$ and $j$, and $n = N / V$ is the density
of the system comprising $N$ particles in a volume $V$.  $\Pi_{ij}$ ($
{} = \Pi_{ji}$) is sampled from a uniform random distribution with
mean and variance $\Pi_0$.  This random component of the momentum
transfer represents the stochastic effect of the collisions and gives
rise to a fluid pressure, while the second, dissipative, term inside
the square brackets of Eqn.~(\ref{OriginalOmega}) is responsible for
the fluid viscosity. The temperature is controlled by $\Pi_{ij}$,
whose variance is a measure of the thermal fluctuations in the system.
Note that $r_c$, which occurs in the factor multiplying the one in
square brackets in Eqn.~(\ref{OriginalOmega}), is a cut-off radius
beyond which no interaction (momentum transfer) is possible. The
presence of this cut-off makes the interactions local and the DPD
algorithm correspondingly fast.

The property of detailed balance is satisfied by
DPD~\cite{bib:espan95} and so a Gibbsian equilibrium state is
guaranteed to exist.  However, the statistical mechanical analysis of
Espa{\~n}ol and Warren~\cite{espanol:dpd} confirmed that the original DPD
model presented by Hoogerbrugge and Koelman~\cite{hoogerbrugge:dpd}
does not lead to the physically correct equilibrium distribution, a
feature which is particularly marked when the relative time step
$\Delta t \ge 1$.

This choice of time step ($\Delta t = 1$) has, nevertheless, been used
in almost all of the DPD work so far reported
on~\cite{hoogerbrugge:dpd,hoogerbrugge:rheoflex,koelman:hard-sphere,schlijper:polymer}.
Two simple modifications to the basic model are suggested by Espa{\~n}ol
and Warren~\cite{espanol:dpd} which ensure that the DPD equilibrium
state is the canonical ensemble.  The first modification is reducing
the time step length (a factor of ten is sufficient), and the second
is the inclusion of an extra factor of $2(1 - r_{ij}/r_c)$ into the
dissipative term in the change of momentum Eqn.~(\ref{OriginalOmega}),
so that the momentum transfer scalar $\Omega_{ij}$ becomes
\begin{equation}\label{AlteredOmega}
 \Omega_{ij} = {3 \left( 1 - {r_{ij} \over r_c} \right) \over \pi
  r_c^2 n} \left[ \Pi_{ij} - 2 \omega \left( 1 - {r_{ij} \over r_c}
  \right) \left( {\mathbf p}_i - {\mathbf p}_j \right) \cdot
  {\mathbf\hat e}_{ij} \right] .
\end{equation}
This alteration guarantees that the model obeys a
fluctuation-dissipation theorem which is very similar to that obtained
in conventional Brownian motion~\cite{espanol:dpd}. The theorem
enables one to relate the amplitude of the noise (that is, the
fluctuations in $\Pi_{ij}$) to the temperature of the system.

In the original paper by Hoogerbrugge and Koelman, it was stated (but
not demonstrated) that the DPD model of a simple one-component fluid
satisfies the Navier-Stokes equations in the mean-field
limit~\cite{hoogerbrugge:dpd}.  Espa{\~n}ol has recently explicitly
derived the hydrodynamic equations for the mass and momentum density
fields in DPD~\cite{bib:espan95}.  However, these equations are not
the central results of Espa{\~n}ol's paper, since the DPD properties of
mass and momentum conservation, coupled with Galilean invariance and
the isotropy of the microscopic equations of motion, effectively
guarantee that at a macroscopic level the Navier-Stokes equations will
emerge.  What is more significant is the correction of the original
simple expressions for the speed of sound and the kinematic viscosity
to more complicated results~\cite{bib:espan95}.

\section{A DPD Model for Binary Immiscible Fluids}
\label{sec:OurMods}

Immiscible fluid mixtures exist because individual molecules attract
similar and repel dissimilar molecules. The most common example of
such behavior arises in mixtures of oil and water; the non-polar,
hydrophobic molecules of oil attract one another through short range
van der Waals forces, while the polar water molecules enjoy more
complex, long-range hydrophilic attractions which are dominated by
electrostatic interactions including hydrogen bonds. At the atomistic
level employed in molecular dynamics, such interactions demand a
detailed treatment. However, to obtain accurate meso- and macroscopic
level descriptions using DPD, the microscopic model can be drastically
simplified. To model the interactions of dissimilar particles in a
binary immiscible fluid within DPD, the simplest modification to the
single phase DPD algorithm is to introduce a new variable, called the
``color'' (by analogy with the Rothman-Keller model), which has two
possible values---for example, ``red'' for oil and ``blue'' for water.
Identical interactions are used between particles of the same color,
while we increase the mean and variance of the random variable
$\Pi_{ij}$ when particles of different color interact, which has the
effect of creating a repulsion between particles from the two
different phases~\cite{hoogerbrugge:rheoflex}.  That is,
\begin{equation}\label{RK-DPD}
\Pi_{ij} \in \cases{
U[0,2\Pi_0] & if particles are of the same phase, \cr
U[0,2(\Pi_0 + \Pi_{rep})] & if particles are of different phases. \cr
}
\end{equation}

It should be emphasized that Eqn.~(\ref{RK-DPD}) is a minimal
modification to the single-phase DPD model and is symmetric under the
interchange of particle colors. It would be quite straightforward to
generalize this to the asymmetric case by, for example, making the
mean $\Pi_0$ of the stochastic terms different for the two colors
and/or likewise adjusting the dissipative terms through the selection
of different values of $\omega$ in Eqn.~(\ref{AlteredOmega}).  In the
present paper, however, we shall not consider such situations,
preferring to concentrate on the properties of the simpler model
implied by Eqn.~(\ref{RK-DPD}).  It has been shown that detailed
balance is also satisfied by the two-phase DPD
model~\cite{coveney:detailed}.  As for the single-phase DPD fluid, one
can show that, macroscopically, the Navier-Stokes equations are obeyed
within homogeneous regions of each of the two immiscible fluids.

The implementation of the DPD algorithm is very similar to that of
conventional MD algorithms~\cite{allen:liquids}.  For example, the
periodic spatial domain (the simulation cell) is divided into a
regular array of equally-sized link cells, such that each side of the
rectangular domain has an integer number of cells and each cell is at
least $r_c$ across.  Each link cell consists of a
dynamically-allocated array of particles and pointers to the
neighboring cells.  Individual particles consist of the
position-momentum vector pair and a color index.

For each time step we calculate first the impulse and then the
propagation step, as described in Eqns.~(\ref{ImpulseStep}) and
(\ref{PropagationStep}).  In the impulse step we iterate through the
particles in each link cell, calculating the change in momentum of
each particle as it interacts with the particles in the same and
neighboring link cells.  Since the momentum is modified by particle
pairs we need to ignore half of the neighboring cells to avoid
duplication.  When considering a new particle pair we first compare
the square of the separation distance with $r_c^2$, and skip to the
next particles if the pair is out of range.  In the propagation step
we iterate through the particles in each link cell again, allowing
them to coast for time $\Delta t$.  The complete state of the system
may be written to file, and other calculations to determine for
example the temperature and pressure of the system can also be
performed.  Given constant $r_c$ and number density $n = N/V$ the
system scales linearly (in both computation time and memory size) with
increasing number of particles, $N$.  To give an example, the
calculation of the motion of a system of 40,000 DPD particles with
number density $n = 4$ for 10,000 time steps takes 11~hours CPU time
and 13~MB memory on an 133~Mflop/s (theoretical) DEC Alpha.

\section{Results}
\subsection{Scaling Laws for Binary Fluid Separation}
\label{sec:BinSep}

Several simulations were run to model the binary separation of a
mixture of two immiscible phases. By definition, for studies of
symmetric quenches, exactly half of the particles were in the ``red''
phase and half in the ``blue'' phase.  Asymmetric quenches were
studied with the ratio of the numbers of colored particles at both
60:40 and 70:30.  The initial state of the system was completely
random; the positions of red and blue particles were chosen from a
uniform distribution, and their velocities were chosen from a uniform
distribution in direction and a Gaussian distribution in magnitude.
Physically, this initial state corresponds to starting with the system
quenched in temperature from a state above the spinodal point at which
the fluids are miscible.

A reasonably large number of particles (40,000) was used in order to
enhance the statistical accuracy of the data obtained during these
simulations.  Care was taken to ensure the box dimensions were
suitable for accurately simulating the phase segregation process for
large times without the size of the domains becoming close to that of
the periodic box, thereby introducing artifacts into the observed
behavior.

\begin{figure}
\caption{Binary Separation Particle Positions}
\label{BinSep:positions}
\begin{center}
\begin{tabular}{lr}
$t = 0\:\delta t$: & $t = 700\:\delta t$: \\
\psfig{file=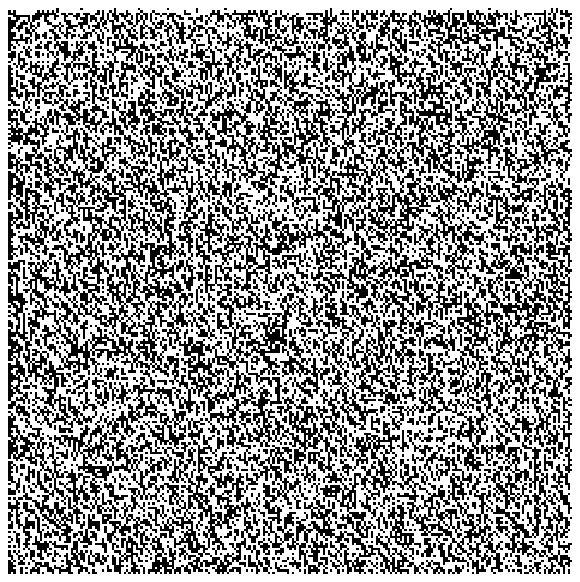} & \psfig{file=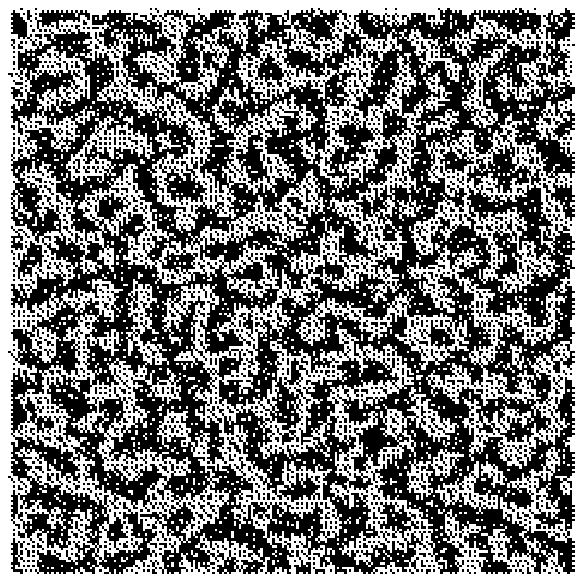} \\
$t = 1300\:\delta t$: & $t = 2500\:\delta t$: \\
\psfig{file=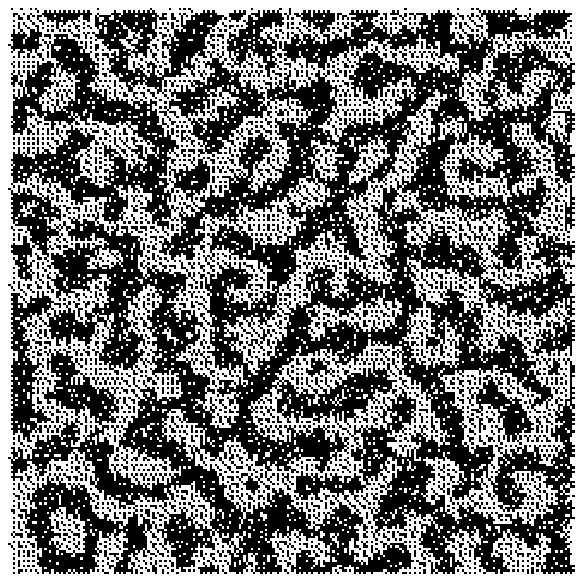} & \psfig{file=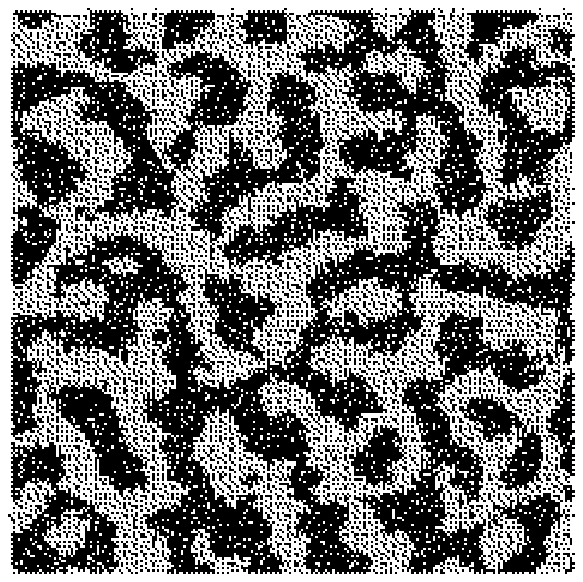} \\
$t = 5000\:\delta t$: & $t = 10000\:\delta t$: \\
\psfig{file=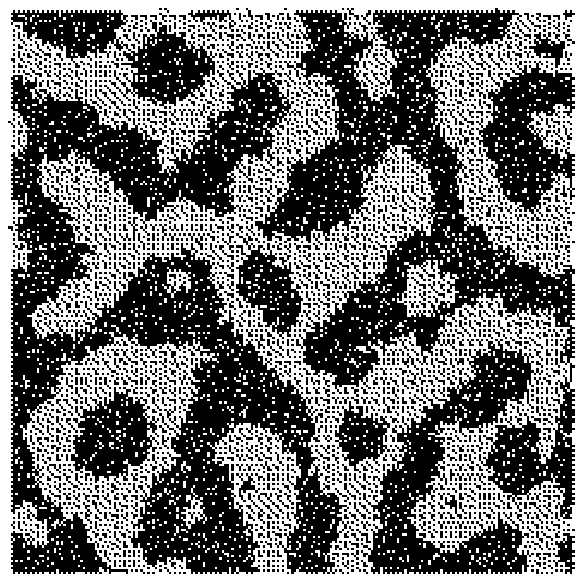} & \psfig{file=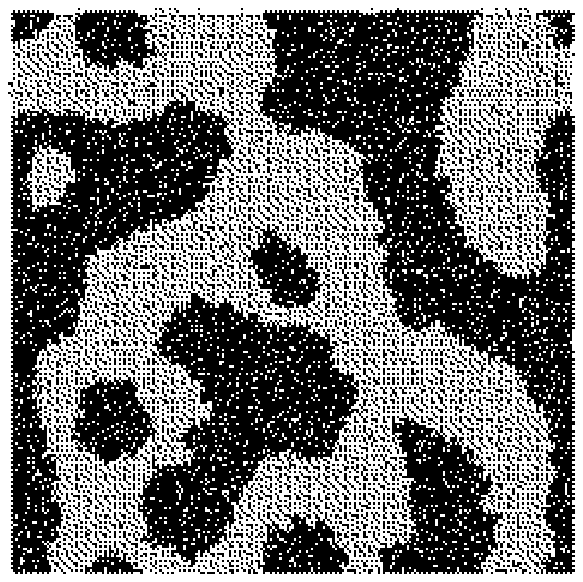} \\
\end{tabular}
\end{center}
\end{figure}

Simulations started from a symmetric quench were allowed to evolve for
10,000 time steps and the asymmetric quenches were evolved for 100,000
time steps.  The state of the system was recorded at regular intervals
throughout.  Fig.~\ref{BinSep:positions} shows the state of a single
system at six different snapshot times following a symmetric quench.
At each time for which the state of the system was recorded, the
structure function
%
\begin{equation} 
S({\mathbf k},t) = \left| {1 \over V} \int {(\rho_b({\mathbf x},t) -
    \rho_r({\mathbf x},t) - \left\langle{\rho_b}\right\rangle +
    \left\langle{\rho_r}\right\rangle ) \exp \left(- 2 \pi i
      \,{{\mathbf k}}\cdot{\mathbf x} \right) } d{\mathbf x} \right|
^2
\end{equation}
%
was calculated.  In this equation the subscripts $b$ and $r$ indicate
blue and red particles, so that $\rho_b({\mathbf x},t)$ is the
(spatial) mass density of the blue particles at time $t$, and
$\left\langle \rho_b \right\rangle$ is the average mass density of
blue particles.  The use of the structure function to characterize
binary fluid phase separation is
widespread~\cite{bib:bal,emerton:domains,rothman:spinodal}. Note that,
because of the imposition of two-dimensional periodic boundary
conditions on the simulation cell, the structure function is only
meaningful when evaluated at points in the Fourier space satisfying
\begin{equation} 
{\mathbf k} = \left( k_x, k_y \right) = \left( {m \over L_x}, 
{n \over L_y} \right); ~~ m,n \in {\mathbb Z}.
\end{equation}
where $L_x$ and $L_y$ are the lengths of the box in the $x$- and
$y$-directions of real space, respectively.

\begin{figure}
\caption{Structure Function and Sample Fit}
\label{BinSep:structure}
\begin{center}
  \vspace{2cm}
  \psfig{file=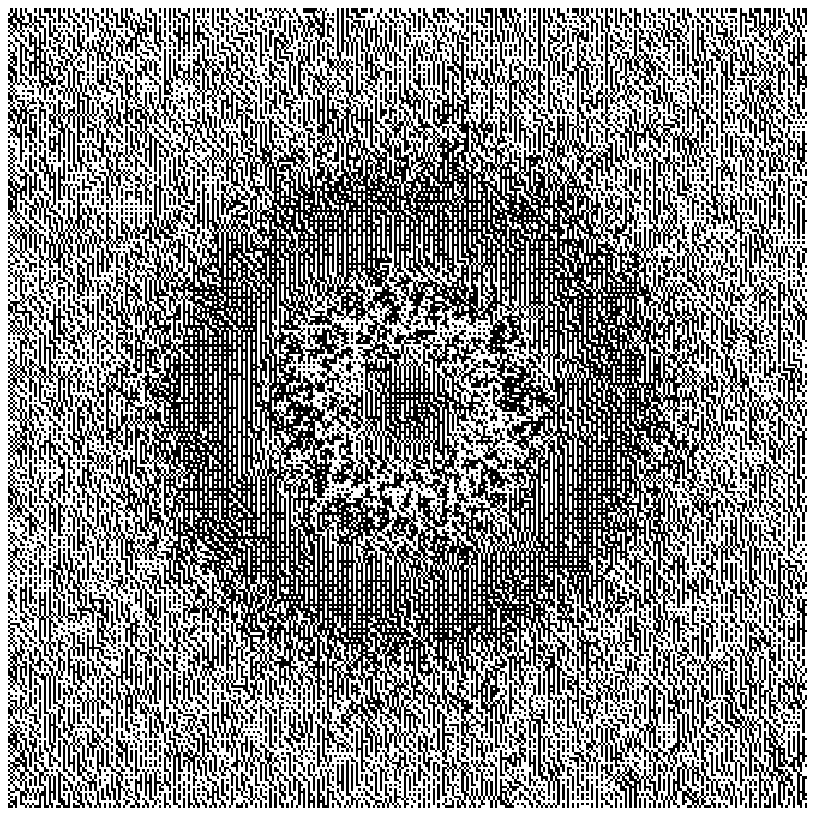} 
  \vspace{2cm} \\ \psfig{file=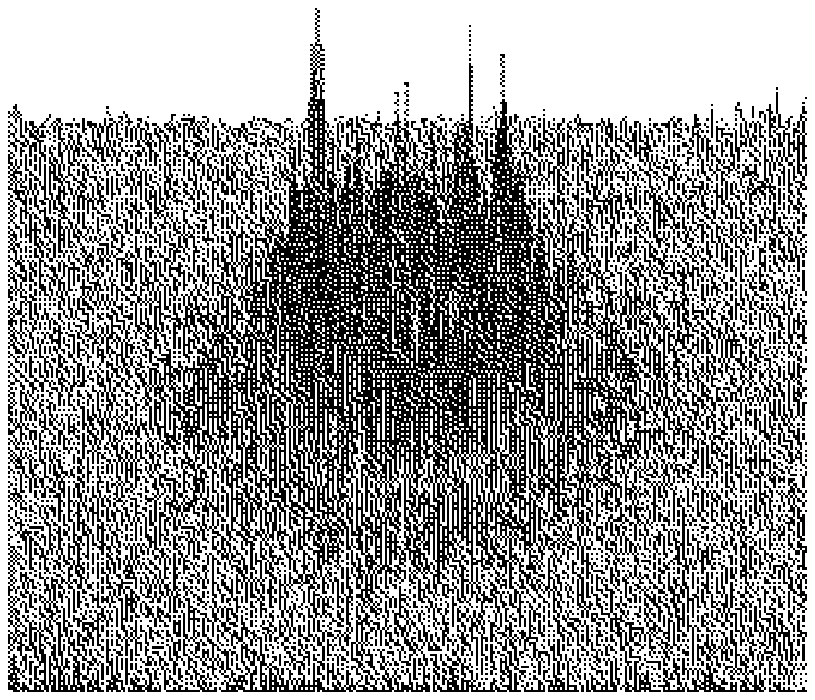}
\end{center}
\end{figure}

To extract the quantity best characterizing the time-dependent domain
size of the state of the system, the mean of $k$ weighted by the
structure function is often calculated.  However, unlike the case of
lattice-based simulation techniques, the positions of the particles in
DPD are described by continuous variables rather than discrete points
on a grid, and so the structure function is meaningful even as
$|{\mathbf k}|$ approaches infinity.  The nature of the system is such
that the structure function has a finite value throughout ${\mathbf
  k}$-space, approaching a constant value at large distances from the
origin, so that the mean is not defined.  Since we are not interested
in the asymptotic value of the structure function, a function
$F({\mathbf k},t)$, which vanishes far from the origin, was fit to the
grid of points containing the meaningful values of $S({\mathbf k},t)$.
The function that we chose as the best fit for $S({\mathbf k},t)$ in
these simulations is
\begin{equation} 
F({\mathbf k},t) = F(|{\mathbf k}|,t) =  c_0 (t) \left| {\mathbf k} 
\right| \exp \left(
  -c_1(t) \left| {\mathbf k} \right|^2 \right).
\end{equation}
Note that $F({\mathbf k},t)$ is taken to have radial symmetry as we
are not interested in the orientation of the domains in real space.  A
sample structure function with its fit $F({\mathbf k},t)$ are shown in
Fig.~\ref{BinSep:structure}.  The reciprocal of the mean of $\left|
  {\mathbf k} \right|$ weighted by $F({\mathbf k},t)$ is denoted
$R(t)$, and is interpreted as the domain size characterizing the state
of the system at time $t$.

\begin{figure}
\caption{Sample $R$ vs. $t$ log-log Plot}
\label{BinSep:loglog}
\begin{center}
  \psfig{file=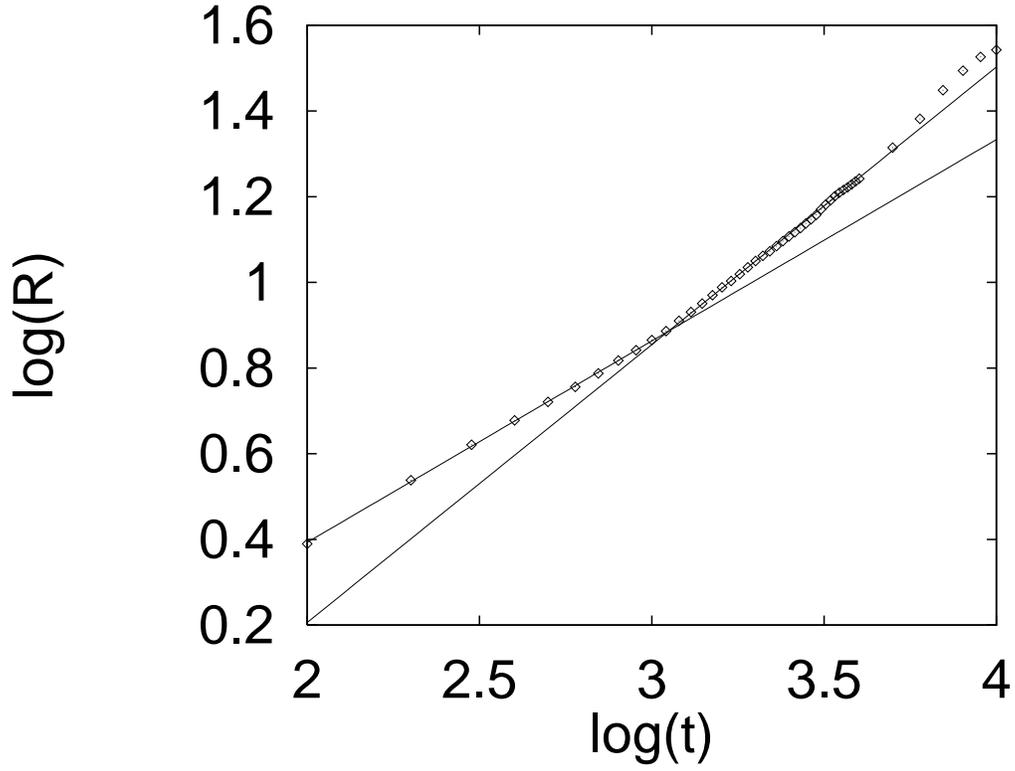,height=4in}
\end{center}
\end{figure}

By plotting our computed values of $R(t)$ on a log-log plot {\em
  versus} $t$, it is easy to see any change in the exponent of the
scaling law
\begin{equation} 
R(t) \propto t^n .
\end{equation}
To obtain accurate results it is important to ensemble-average several
simulations differing in their random initial conditions.  For each
simulation, the log-log plot was examined for possible exponents.
From each of the symmetric quenches it was possible to mark the
existence of two separate scaling regimes, with the crossover at
roughly $t = 1000 \Delta t$.  Only a single scaling regime was
observed from the asymmetric quenches, even though the simulations
were allowed to evolve an order of magnitude longer than the symmetric
quenches and the final domains were significantly larger.  For each
regime in all simulations a straight line was fit to the data to
determine the exponent.  The complete results for symmetric quenches
are listed in Table~\ref{BinSep:exponents}, while a sample log-log
plot is shown in Fig.~\ref{BinSep:loglog}.  All of the figures in this
section are derived from one and the same simulation, and so may be
directly compared.  The results in Table~\ref{BinSep:exponents}
strongly suggest a crossover from $n = \frac{1}{2}$ to $n =
\frac{2}{3}$, which is in agreement with results from lattice gas
automata~\cite{emerton:domains}, molecular dynamics~\cite{bib:vat} and
Langevin dynamics simulations~\cite{bib:lwac} as well as a
renormalization group analysis~\cite{bib:bray}.  A growth exponent of
$n = \frac{1}{2}$ was observed from the asymmetric quenches,
independent of the ratio of particle numbers.  This exponent was
observed at all times in these off-critical quenches, just as in
previously reported Langevin simulations~\cite{bib:walc,bib:lwac}.

Some short simulations (3,000 time steps long) were also run to
determine the early-time exponent for a symmetric quench where
momentum is not conserved.  The momentum transferred between a pair of
interacting particles was violated by adding a vector 15--20\% of the
magnitude of the interaction and random in direction (chosen from a
uniform distribution) to each particle's momentum.  The growth
exponent observed in these simulations was $n = 0.329 \pm 0.005$, in
agreement with theory~\cite{bib:bray}.

The complete set of model parameters used in these simulations is
listed in Table~\ref{BinSep:parameters}.  In this table, $L_x$ and
$L_y$ are the box dimensions, while $n$ is the number density $N/V$,
where $V$ is the two-dimensional volume of the box. The value of
$\Pi_0 = 5.0$ was chosen since the immiscible phases then regularly
reach complete phase separation asymptotically and yet the system does
not behave too randomly.

\subsection{Bubble Surface Tension}
\label{sec:BubTen}

The detailed way in which phase separation occurs in binary fluids
depends, among other things, on the interfacial tension which exists
between the two immiscible phases. In particular, as noted earlier,
when hydrodynamic effects are important, the crossover between
diffusive (Lifshitz-Slyozov) and hydrodynamic regimes should occur
when the domain size $R(t)$ is larger than the hydrodynamic length
$R_{h} = {\nu}^2 / \rho \sigma$, where $\nu$ is the kinematic
viscosity, $\rho$ is the fluid density and $\sigma$ is the surface
tension coefficient.  A further important test of our DPD model for
binary fluid separation is thus to check on the existence of a surface
tension between the two phases by confirming the validity of Laplace's
law using a series of bubble simulations inspired by earlier lattice
gas analogues~\cite{emerton:domains,adler:tension}.

\begin{figure}
\caption{Sample Tension Bubble at Equilibrium ($R_0 = 0.125 L_x$)}
\label{BubTen:SampleBubble}
\begin{center}
  \psfig{file=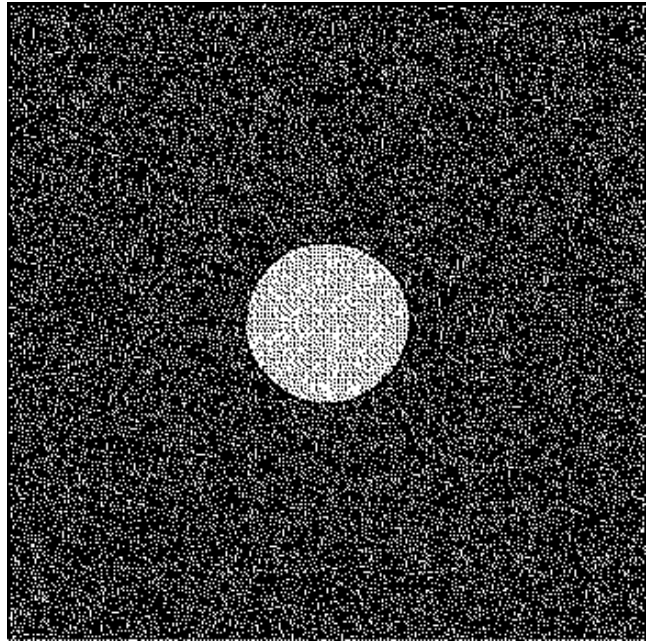}
\end{center}
\end{figure}

As with the domain growth simulations of Sec.~\ref{sec:BinSep}, 40,000
particles were placed in a two-dimensional periodic box of the same
dimensions as those listed in Table~\ref{BinSep:parameters}, but the
simulations were now run for 40,000 time steps.  (For a few of the
smaller bubbles, simulations were run with only 6,400 particles and to
5,000 time steps. For these smaller bubbles accurate results could be
obtained without the additional computation required by the larger
system.)  The results here have the same model parameters as in
Sec.~\ref{sec:BinSep} (see Table~\ref{BinSep:parameters}) with the
sole exception of $\Pi_0$, which was set to 0.25 in order to increase
the signal-to-noise ratio.  The initial state of the system has all
the particles placed with random positions and velocities, but the
particles are red within a radius $R_0$ of the center of the box, and
blue outside this distance.  This bubble changes rapidly within the
first few time steps, but settles down to a state approximating
equilibrium after about 8,000 steps for the larger simulations (2,000
time steps for the smaller simulations).
Fig.~\ref{BubTen:SampleBubble} shows one such bubble at equilibrium.

These bubble experiments thus enable us to compute the {\em
equilibrium} value of the interfacial tension between red and blue
phases. Equilibrium statistical mechanics tells us that we can write
the instantaneous pressure function (${\mathcal P}$) of a system in
terms of the internal virial (${\mathcal W}$) and the instantaneous
temperature function (${\mathcal T}$)~\cite{allen:liquids} as
\begin{equation} 
{\mathcal P} = {N k_B {\mathcal T} + {\mathcal W} \over V} , 
\end{equation}
where
\begin{equation} 
N k_B {\mathcal T} 
= {1 \over 3} \sum_i { \left| {\mathbf p}_i \right|^2  \over m_i} ,
\end{equation}
and
\begin{equation} 
{\mathcal W} = {1 \over 3} \sum_i \sum_{j>i} {\mathbf r}_{ij} \cdot
{\mathbf\hat e}_{ij} \Omega_{ij} ,
\end{equation}
where $k_B$ is the Boltzmann constant.  The thermodynamic pressure
($P$) and temperature ($T$) of the system are the time averages of
${\mathcal P}$ and ${\mathcal T}$, respectively.  It is possible to
compute a temperature because the important property of detailed
balance is satisfied by DPD~\cite{bib:espan95}; the fluctuation
dissipation theorem relates this to the noise. (We note in passing
that Hoogerbrugge and Koelman~\cite{hoogerbrugge:dpd} gave an equation
of state for a homogeneous DPD fluid which relates the fluid pressure
to the fluid density.  However, since the theoretical basis for this
equation is not adequately explained we have preferred to calculate
the pressure using statistical mechanical first principles.)

\begin{figure}
\caption{Pressure Difference vs. 1/R}
\label{Pdiff_vs_R}
\begin{center}
  \psfig{file=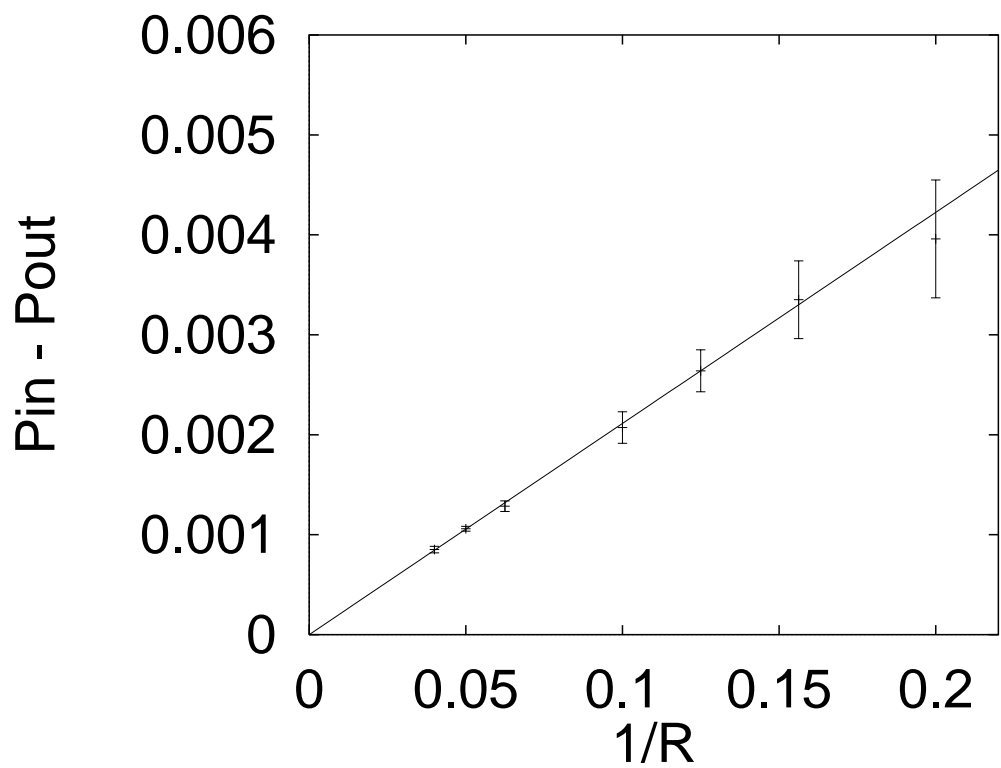,height=4in}
\end{center}
\end{figure}

By considering only the blue particles far from the interface ($r >
1.3 R_0$) we can calculate the thermodynamic pressure in homogeneous
regions outside the bubble ($P_{out}$).  Similarly, we can calculate
the pressure inside the bubble ($P_{in}$) by using particles with $r <
0.7 R_0$.  For a circular bubble in two dimensions, Laplace's
law~\cite{adler:tension} says
\begin{equation} 
P_{in} - P_{out} = {\sigma \over R_0} .
\end{equation}
To verify this we calculate the mean pressure difference at several
values of $R_0$, and plot the pressure difference $P_{in} - P_{out}$
versus the reciprocal of $R_0$.  For each of the simulations the
equilibrium pressure difference was averaged to a single mean value.
The mean pressure difference and corresponding error for each bubble
size reported in Fig.~\ref{Pdiff_vs_R} are the mean and standard error
of the mean for the set of simulations at each bubble size.  We can
see from our results in Fig.~\ref{Pdiff_vs_R} that we have the
expected linear behavior.  The slope of this line is the surface
tension coefficient, $\sigma$.

\section{Conclusions}
\label{sec:Conclu}

We have studied binary immiscible fluid behavior in two spatial
dimensions using a novel simulation technique called dissipative
particle dynamics. We have found algebraic scaling laws in agreement
with expectations~\cite{bib:bray}, the two-dimensional growth
exponents being $\frac{1}{2}$ and $\frac{2}{3}$ at early and late
times respectively for symmetric quenches, and $\frac{1}{2}$
throughout for asymmetric quenches.  Symmetric quenches for which
momentum transfer was violated displayed an early-time growth exponent
of $\frac{1}{3}$.  This scaling behavior has previously been observed
in molecular dynamics~\cite{bib:vat}, Langevin
dynamics~\cite{bib:lwac} and lattice gas
automata~\cite{emerton:domains} simulations, and is also in accord
with the results of a renormalization-group approach which takes into
account the mechanism of droplet coalescence due to noise-induced
Brownian motion~\cite{bib:bray}.  We have also verified that Laplace's
law holds in a series of simple bubble experiments, confirming the
existence of a surface tension between the two phases.

We conclude that dissipative particle dynamics is a promising method
for simulating the properties of fluid systems.

\section*{Acknowledgments}

We are grateful to Bruce Boghosian, Alan Bray, John Melrose and Pep
Espa{\~n}ol for helpful discussions.  KEN gratefully acknowledges
financial support from NSERC (Canada) and the United Kingdom ORS
Awards Scheme.



\newpage

\begin{table}
\caption{Binary Separation Exponents for Symmetric Quenches}
\label{BinSep:exponents}
\begin{tabular}{ddddd}
Simulation & Slope 1  & Slope 2 & $\log t$  & $\log R$  \\
number     &          &         & crossover & crossover  \\
\tableline
 8         &  0.4701  &  0.648  &  3.000    &  0.865  \\
 9         &  0.4648  &  0.640  &  3.000    &  0.869  \\
11         &  0.4805  &  0.639  &  3.000    &  0.886  \\
12         &  0.4788  &  0.642  &  3.114    &  0.932  \\
13         &  0.4804  &  0.649  &  3.079    &  0.916  \\
14         &  0.4859  &  0.627  &  3.000    &  0.884  \\
15         &  0.4759  &  0.676  &  3.114    &  0.929  \\
16         &  0.4717  &  0.673  &  3.079    &  0.912  \\
17         &  0.4855  &  0.696  &  3.114    &  0.943  \\
\tableline
Mean       &  0.477   &  0.65   &  3.06     &  0.90  \\
StdDev     &  0.007   &  0.02   &  0.05     &  0.03  \\
\end{tabular}
\end{table}

\begin{table}
\caption{Binary Separation Model Parameters}
\label{BinSep:parameters}
\begin{tabular}{cd}
Model           &               \\
Parameter       & Value         \\
\tableline
$\Delta t$      & 0.1           \\
$m$             & 1.0           \\
$n$             & 4.0           \\
$N$             & 40000.        \\
$\omega$        & 2.0           \\
$\Pi_0$         & 5.0           \\
$\Pi_{rep}$     & 0.3           \\
$r_c$           & 1.3           \\
$L_x$       & 100.0         \\
$L_y$       & 100.0         \\
\end{tabular}
\end{table}

\end{document}